\documentclass[conference]{IEEEtran}
\usepackage{dcolumn}% Align table columns on decimal point
\usepackage{bm}% bold math
\usepackage{graphicx}
\usepackage{amssymb}
\usepackage{amsmath}   % From the American Mathematical Society
\begin{document}

\title{Frequency-Tunable Josephson Junction Resonator for Quantum Computing}

\author{\authorblockN{K. D. Osborn, J. A. Strong, A. J. Sirois, and R. W. Simmonds}
\authorblockA{National Institute of Standards and Technology (NIST)\\
Boulder, Colorado 80305\\
Email: osborn@boulder.nist.gov}}

% avoiding spaces at the end of the author lines is not a problem with
% conference papers because we don't use \thanks or \IEEEmembership

%Homer Simpson\authorrefmark{2},
%Montgomery Scott\authorrefmark{3} and
%Eldon Tyrell\authorrefmark{4}}
%Telephone: (800) 555--1212, Fax: (888) 555--1212}
%\authorblockA{\authorrefmark{4}Tyrell Inc., 123 Replicant Street, Los Angeles, California 90210--4321}}

%\specialpapernotice{(Invited Paper)}

\maketitle

\begin{abstract}
We have fabricated and measured a high-Q Josephson junction resonator with a tunable resonance frequency. A dc magnetic flux allows the resonance frequency to be changed by over 10 \%. Weak coupling to the environment allows a quality factor of $\thicksim$7000 when on average less than one photon is stored in the resonator. At large photon numbers, the nonlinearity of the Josephson junction creates two stable oscillation states. This resonator can be used as a tool for investigating the quality of Josephson junctions in qubits below the single photon limit, and can be used as a microwave qubit readout at high photon numbers.

\end{abstract}


% for peerreview papers, inserts a page break and creates the second title.
% Will be ignored for other modes.
\IEEEpeerreviewmaketitle

\section{Introduction}
% no \PARstart

Josephson junction resonators are important elements in superconducting quantum computing circuits.  For example, a phase qubit is a Josephson junction (JJ) resonator in which the two lowest energy levels of the resonator describe the qubit states\cite{Martinis}. In addition, resonators are useful in measuring qubits in dispersive \cite{SchoelkopfOsc+Qubit} and latching \cite{DevoretOsc+Qubit} readouts.  

We have developed a Josephson junction resonator that exhibits a bifurcation of states similar to that in reference \cite{DevoretOsc+Qubit}, but has notable differences.  The design of the resonator looks similar to a phase qubit in that the Josephson junction is in a superconducting loop. This allows flux bias into the loop to significantly vary the resonance frequency.  As we discuss below, this has important implications for a qubit readout. In addition, our resonator is weakly coupled to the environment, which allows us to probe the quality of a single Josephson junction when on average there is less than one photon in the resonator.  Phase qubit measurements indicate that the quality of a Josephson junction for quantum computing is hindered by two level systems within\cite{Cooper}\cite{DielectricLoss}. 

\section{Circuit Theory}
The nonlinear resonator circuit that we discuss in this paper is shown within the dashed box in Figure 1.  The JJ with capacitance $C_\textrm{J}$ and inductance $L_\textrm{J}$ is in series with an inductance $L_\textrm{S}$.  Parallel to this is an inductance $L_\textrm{P}$ and a capacitor $C_\textrm{P}$. $L_\textrm{P}$ allows us to apply a dc bias current across the JJ with an applied flux $\Phi_\textrm{a}$ and $C_\textrm{P}$ is used to limit the resonance frequency.  This resonator is coupled to transmission lines through the capacitors $C_\textrm{C}$ (outside the dashed box).  Without coupling, the resonator has a quality factor $Q_\textrm{0}$, which is linear with the effective parasitic rf resistance $R_\textrm{0}$.  With coupling, the resonator is loaded to the measured quality factor $Q_\textrm{l}=1/(1/Q_\textrm{0}+1/Q_\textrm{e})$, where $Q_\textrm{e}$ is the external quality factor, which depends on $C_\textrm{C}$.  One transmission line carries an incoming voltage wave of rms amplitude $V_\textrm{in, rms}$ with input power $P_\textrm{in}=V_\textrm{in rms}^2/Z_0$.  Similarly, the other transmission line carries the output (transmitted) voltage wave of rms amplitude $V_\textrm{out, rms}$, with an output power $P_\textrm{out}=V_\textrm{out, rms}^2/Z_0$.  

The ac voltage $V$ from the supercurrent $I$ is $V=\gamma\dot{I}/\sqrt{I_\textrm{c}^2-I^2}$, where $\gamma=\Phi_0/2 \pi=h/2e$.  The supercurrent through the junction $I=I_\textrm{dc}+I_\textrm{ac}$ can be written in terms of dc and ac components.  For small currents $I$, where $|I|<<I_\textrm{c}$, we can write $V/\dot{I} = f_\textrm{dc}+f_\textrm{ac1}I_{ac}+f_{ac2}I_\textrm{ac}^2+...$, where $f_\textrm{dc}=\gamma/\sqrt{I_\textrm{c}^2-I_\textrm{dc}^2}$. Assuming that the ac current through the device is $i_\textrm{ac}\sin(\omega t)$ and including only the first harmonic in $\omega$, a power expansion yields $V = i_\textrm{ac}\omega\cos(\omega t)L_\textrm{J}$, where the constant nonlinear inductance is $L_\textrm{J}=(f_\textrm{dc}+f_\textrm{ac2}i_\textrm{ac}^2/4)$ and  $f_\textrm{ac2}=\frac{\gamma}{2(I_\textrm{c}^2-I_\textrm{dc}^2)^{3/2}}(1+\frac{3 I_\textrm{dc}^2}{I_\textrm{c}^2-I_\textrm{dc}^2})$.  The input voltage wave with the coupling capacitor creates an ac current bias across the resonator circuit, and with the inductance $L_J(i_{ac})$, we calculate the voltage in the resonator and the voltage transmitted though the resonator.  Like a particle with a nonlinear restoring force, near resonance the system exhibits two stable oscillation states for the same drive amplitude.  We have found agreement between this analysis and a fourth-order Runge-Kutta simulation, which includes the full shape of the potential.

%  At zero flux bias and at the linear resonant frequency, the total current of the resonator can be expanded to find $V_\textrm{in rms}/V_\textrm{out rms}=\sqrt{a+b V_\textrm{out rms}^4}$ for small $V_\textrm{out rms}$,  where $a$ and $b$ are functions of the device parameters.  This implies that $V_\textrm{in rms}/V_\textrm{out rms}$ should be nearly constant at low values of $V_\textrm{out rms}^2$ and be proportional to $V_\textrm{out rms}^2$ for larger values of $V_\textrm{out rms}$.

\begin{figure}
\centering
\includegraphics[width=3.2in]{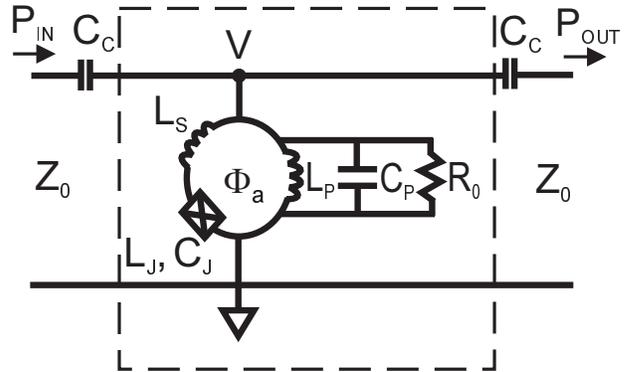}
\caption{Resonator Circuit: Transmission lines of impedance $Z_0$ connect to the resonator through the coupling capacitors $C_\textrm{c}$.  The resonator contains a parasitic effective internal resistance $R_{0}$.}
\label{fig_sim}
\end{figure}

\section{Device Design}

We fabricated the device using conventional photolithography, as shown in figure 2.  Between the two coupling capacitors $C_{c}$ the resonator contains several lumped elements.  Two large interdigital capacitors were used to form $C_\textrm{p}$.  Al wires were used to form $L_{p}$ and $L_\textrm{s}$, which have nominal values of 400 pH and 120 pH, respectively.  Two small holes in a $\textrm{SiN}_\textrm{x}$ wiring insulator define two parallel JJs with a total area of 11 $\mu \textrm{m}^2$, which we refer to as a single junction in the rest of this paper.  On the same wiring layer is a via, connecting the other end of the JJ to the remainder of $L_S$.  The devices were measured at $T\thickapprox$ 45 mK by use of a HEMT amplifier at $T=4$ K to amplify the power out of the device.

The loss of the components of the resonator was designed to be low, so that we will be sensitive to any residual losses from within the JJ producing a $Q_\textrm{0}$ of $\lesssim 10^{4}$.  The loss tangent of the interdigital capacitors is negligible, since it was fabricated on a sapphire substrate with a loss tangent of $\leqq 10^{-6}$. $\textrm{SiN}_\textrm{x}$, with a moderately good loss tangent of $\thickapprox0.0003$, near the JJ produces a comparatively small capacitance ($\thicksim$30 fF) which contributes negligible loss. Finally, in devices with and without a JJ we measured samples with a medium to high external quality factor $Q_\textrm{e}>Q_\textrm{0}/5$ to accurately extract $Q_\textrm{0}$. 

\begin{figure}
%\centering
\includegraphics[width=3.2in]{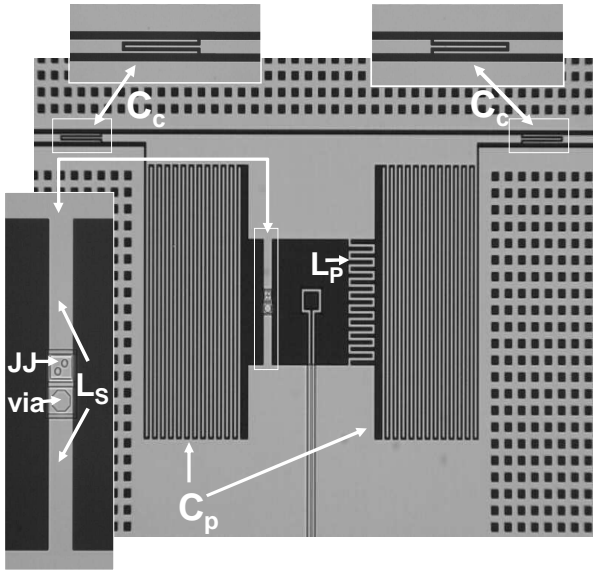}
\caption{Micrograph of the fabricated resonator circuit.  Metallic regions appear in gray.}
\label{device}
\end{figure}

\section{Results}

In figure 3 we show the resonance frequency as a function of flux bias $f_r(\Phi_a)$, taken from a transmission measurement of the resonator.  As expected, we observed periodic behavior of the resonant frequency with the flux quantum. At zero flux bias the inductance of the junction $L_J$, was a minimum, which in turn creates a maximum in the resonance frequency at f=8.2423 Ghz. Moving away from zero flux bias we observed a decrease in resonance frequency which continues below 7.1 GHz, giving us an observed change in resonance frequency of over $10\%$.   Note that this device exhibits a large slope $df_r/d\Phi_a$ away from zero flux bias.
\begin{figure}
\centering
\includegraphics[width=3.2in]{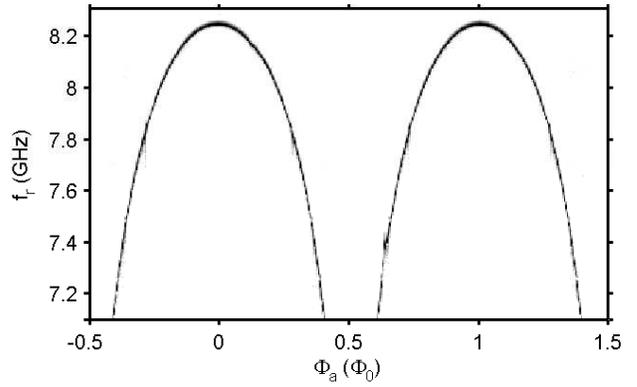}
\caption{Measured resonance frequency as a function of flux bias.}
\label{spec}
\end{figure}

In figure 4 we show the frequency sweeps for different input powers at zero flux bias.  At the lowest input power shown we have less than one photon in the cavity and the output amplitude is proportional to the input power.  As we increase the power the peak becomes asymmetric.  At the highest power shown, two stable oscillation states are reproducibly observed as the frequency is swept in two opposite directions.  Thus the resonator has two stable oscillation states even though it resides in a single potential well.  Because the resonant frequency changes with flux bias, the switching frequency between the two stable oscillation states is also sensitive to the flux bias. As a result this device can be used as a low-loss flux detector, such as a qubit readout.

\begin{figure}
\centering
\includegraphics[width=3.2in]{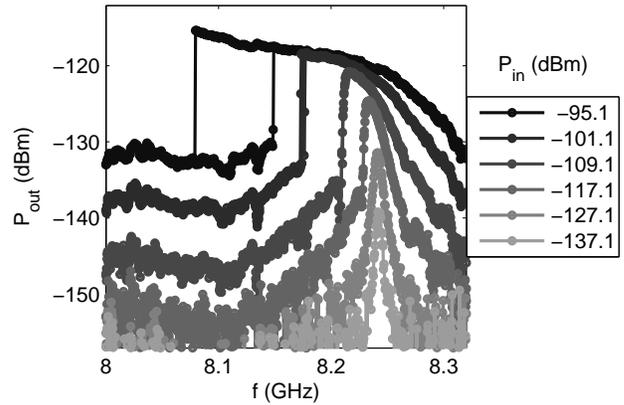}
\caption{Power out of the resonator as a function of frequency for different input powers at zero flux bias.}
\label{freq}
\end{figure}

The fit parameters for figures 3 and 4 have been simultaneously compared to theory curves in order to find the unknown parameters of the circuit.  In the fit presented here, we fixed $L_\textrm{p}$ and $L_\textrm{s}$ to the nominal values (above) and we set $C_J = 0$ pF.  We have also fit the data with the value $C_\textrm{J}=0.2$ pF, in which case the quality of the fit is also good.  From qubit data, we estimate that $C_\textrm{J}\thickapprox0.4$ pF, which is larger than the best fit values, but still small compared to the main capacitor $C_\textrm{p}$.

A fit to the spectroscopy in figure 3 (not shown) was used to find $I_\textrm{c}=0.905$ $\mu$A ($L_\textrm{J0}=364$ pH) and $C_P$=1.68 pF.  A Lorentzian fit (not shown) to the lowest power frequency sweep in figure 4, where the response is linear, was used to find $C_c=12.5$ fF and $R_{0}=82,000$ $\Omega$, which gives $Q_{0}$ = 7,100.  For reference, we measured a device that was the same except that it did not have a JJ or inductor $L_S$ (and also had a different $C_c$).  This reference device had a $Q_{0}\thickapprox10^5$, which is much larger than the device without a JJ, indicating that additional loss may be caused by the junction. The values of $Q_{0}$ in these devices also show that this technique is appropriate for studying the loss in a Josephson junction.  Below we show that the JJ resonator described above is also measured below the single-photon level.

The circles in figure 5 show the data for a power sweep taken at the fixed frequency corresponding to the low power resonance frequency for zero flux bias (f = 8.2423 Ghz).    At low voltages, $V_\textrm{in, rms}/V_\textrm{out, rms}$ is constant in $V_\textrm{out, rms}^2$, indicating linear response.    Linear response continues to $V_\textrm{out, rms}^2\thickapprox0.002$ $\mu\textrm{V}^2$, where the supercurrent amplitude is $i_{ac}$ = 0.0861 $I_{c}$. At larger $V_\textrm{out, rms}$, $V_\textrm{in rms}/V_\textrm{out, rms}$ is linear in $V_\textrm{out, rms}^2$, reflecting the expected nonlinearity in $L_J$ with ac drive amplitude.

In figure 5, the fit parameters for low power data in figures 3 and 4 were used to generate the theoretical curve for the power sweep.   These fit parameters and our model describe the entire range of powers shown in figure 5. The agreement between data and the theoretical curve indicates that our fit values and model are reasonably accurate.  

In order to analyze the quality of Josephson junctions for use in a future qubit, one would like to perform the analysis at (on average) less than one photon in the resonator.  On resonance the cavity stores an energy $E_{cav}=\langle n \rangle hf$ for an average number of $n$ photons $\langle n \rangle$. In our symmetric resonator photons exit via the output cable at a rate $\pi f/Q_{e}$, yielding a power $P_{out} = \langle n \rangle\pi h f^2/Q_\textrm{e}$. For our device, $Q_{e}=1,900$ indicating that we have less than one photon on average in the cavity when $P_\textrm{out}<-131.3$ dBm and $V_\textrm{out, rms}^2<0.0037$ $\mu\textrm{V}^2$.  Since we analyze the quality of the single JJ resonant circuit below this power, where the response is linear, we are in the appropriate power regime for investigating JJs used in qubits.

\begin{figure}
\centering
\includegraphics[width=3.2in]{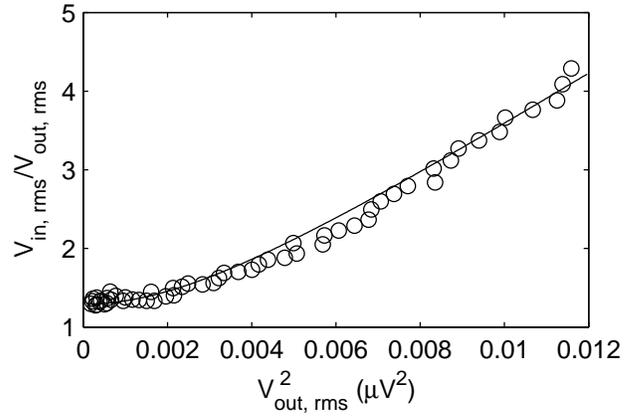}
\caption{$V_\textrm{in, rms}/V_\textrm{out, rms}$ versus $V_\textrm{out, rms}^2$ at $\Phi_\textrm{a}=0$ and the low power resonance frequency for zero flux bias (f=8.2423Ghz).  The circles show the experimental data, and the line shows the expected theoretical result for parameters determined by fits to figure 3 and figure 4.}
\label{power}
\end{figure}

\section{Conclusion}

We have designed and measured a high-Q Josephson junction resonator with a tunable resonance frequency.  We are able to describe the flux bias and nonlinear ac data in terms of a simple lumped element model with a nonlinear inductance for the junction. At low powers, with less than one photon in the resonator cavity, we observe a quality factor of $Q_{0} = 7,100$. This high quality factor below the single photon limit and the performance relative to a reference device, shows that the resonator is appropriate for studying loss in a JJ for qubits. At high powers we observe two stable oscillation states with reproducible switching points, and therefore this device may be useful as a future qubit readout.


\section*{Acknowledgment}
% optional entry into table of contents (if used)
%\addcontentsline{toc}{section}{Acknowledgment}
We acknowledge helpful discussions with R. Kautz. This work was funded by NIST's Quantum Information Program and in part through DTO Grant MOD110708.  Publications of NIST, a government agency, are not subject to US copyright.



% http://www.ctan.org/tex-archive/macros/latex/contrib/supported/IEEEtran/bibtex
%\bibliographystyle{IEEEtran.bst}

\end{document}